# Controllable Asymmetric Matter-wave Beam Splitter and Ring Potential on an Atom Chip

S. J. Kim,[1] H. Yu,[1,2] S. T. Gang,[1] D. Anderson[2], and J. B. Kim[1]
[1]Department of Physics Education, Korea National University of Education, Chung-Buk 363-791, Republic of Korea
E-mail address: jbkim@knue.ac.kr
[2]Department of Physics and JILA, University of Colorado, Boulder, Colorado 80309-0440, USA

We have constructed an asymmetric matter-wave beam splitter and a ring potential on an atom chip with Bose-Einstein condensates using radio-frequency dressing. By applying rf-field parallel to the quantization axis in the vicinity of the static trap minima added to perpendicular rf-fields, versatile controllability on the potentials is realized. Asymmetry of the rf-induced double well is manipulated without discernible displacement of the each well along horizontal and vertical direction. Formation of an isotropic ring potential on an atom chip is achieved by compensating the gradient due to gravity and inhomogeneous coupling strength. In addition, position and rotation velocity of a BEC along the ring geometry are controlled by the relative phase and the frequency difference between the rf-fields, respectively.

**PACS number(s):** 03.75.Be, 37.10.Gh, 67.85.De

Microscopic magnetic traps utilizing an atom chip are of interest for the study of neutral atoms and Bose-Einstein condensates (BECs) since they offer a degree of design flexibility, precise control over the trapped atoms, and feasibility of integration of atom-optical elements [1]. Atom chips enable the straightforward application of radio frequency (rf) radiation to atoms in an otherwise static magnetic trap for modifying its potential [2-5]. Adiabatic rf-dressed potentials have the benefit of being smooth and offer long phase coherence times while maintaining high confinement during flexible deformation of the trap geometry [6]. The most common geometries for rf-dressed adiabatic potentials are double-wells and ring potentials [6, 7]. The rf-induced double well acts as a matter-wave beam splitter and thus has been used for atom interferometry [8, 9] utilized for the study of precision measurement [10, 11], phase dynamics [12-15], and squeezed and entangled states [16-18], etc. In addition, the double-well geometry was used for the study of superfluidity [19]. Precise control over asymmetry of a double well is very important since the contrast and the coherence time of the interferometer depend on the balance of the two wells [8, 17]. Trap asymmetry can be used as a means to shift the relative phase of interferometer arms [4], and as a means of setting the bias of a Josephson junction for superfluidity studies [19-21]. At the same time, ring-shaped traps have been studied by many groups [22-30], since the geometry can be used to construct to multiply-connected systems with behavior not found in other types of trap [27, 31, 32]. Such ring traps also have practical potential for Sagnac interferometry for inertial sensing [6, 27] and are also of fundamental interest as an analog of the superconducting quantum interference device (SQUID) [28, 33], and for the study of superfluidity and quantization of the angular momentum [23, 29, 30, 34, 35], etc. Even though a ring potential on an atom chip was suggested in Ref. [6], it had not been realized due to gravity and inhomogeneous coupling strength of the rf-field [7].

In this letter, we report on the realization of the controllable asymmetric matter-wave beam splitter and a ring potential on an atom chip. The theory of the asymmetric rf-induced potentials is discussed and the formation of various geometries is explained. Experimental results showing the versatility of the rf-induced potentials are presented. The fractional atom number distribution is used as a measure of the double-well asymmetry. For a ring potential, homogeneity was obtained for the first time by adjusting the asymmetry to compensate the inhomogeneity. In addition, a rotation scheme to introduce angular momentum into a ring BEC is also investigated.

Since the coupling between a static field and rf-fields critically depends on the vector characteristics of the fields, controlling the vector character is the key to the versatility of the rf-induced potentials [6]. For this, we create the rf-dressed potentials by combining a static Ioffe magnetic trap $\boldsymbol{B}_S(\boldsymbol{r})$ with a three-components rf-field $\boldsymbol{B}_{rf}$ having frequency $\omega_{rf}$ and relative phase shifts, $\delta$ and $\delta_z$.

$$\boldsymbol{B}_S(\boldsymbol{r}) = Gx\boldsymbol{e}_x - Gy\boldsymbol{e}_y + B_I\boldsymbol{e}_z \quad (1)$$

$$\boldsymbol{B}_{rf} = B_{rf}^A \boldsymbol{e}_x \cos(\omega_{rf} t) + B_{rf}^B \boldsymbol{e}_y \cos(\omega_{rf} t - \delta)$$
$$+ B_{rf}^z \boldsymbol{e}_z \cos(\omega_{rf} t - \delta_z) \quad (2)$$

Here $G$ is the gradient of the static trap, $B_I$ is the magnitude of the Ioffe field, $B_{rf}^A$, $B_{rf}^B$ and $B_{rf}^z$ are the amplitudes of the rf-fields and $t$ is time. Following [6], the corresponding rf-induced potentials can be written as

$$V_{ad}(\mathbf{r}) = m_F g_F \mu_B \sqrt{\Delta(\mathbf{r})^2 + \Omega(\mathbf{r})^2} \quad (3)$$

, where $m_F$ is the magnetic quantum number, $g_F$ is the g-factor of the hyperfine state, and $\mu_B$ is Bohr's magneton. From the Rabi frequency $\Omega(\mathbf{r})^2$, an additional asymmetry component on top of the symmetric potential without $B_{rf}^z$ can be found as [36]:

$$\frac{GB_{rf}}{2|\mathbf{B}_S(\mathbf{r})|^2} B_{rf}^z \Big[ \sin\delta_z \big( |\mathbf{B}_S(\mathbf{r})| y + B_I y \sin\delta + |\mathbf{B}_S(\mathbf{r})| x \cos\delta \big)$$
$$- \cos\delta_z \big( B_I x - B_I y \cos\delta + |\mathbf{B}_S(\mathbf{r})| x \sin\delta \big) \Big]. \quad (4)$$

For double well potentials, the asymmetry term is reduced when $\delta_z = 0$ as,

$$\frac{-GB_I B_{rf}}{2|\mathbf{B}_S(\mathbf{r})|^2} B_{rf}^z (x \mp y); \quad (5)$$

- and + signs indicate $\delta = 0$ and $\pi$, respectively. For a ring potential, $\delta = \pi/2$, the asymmetry term can be denoted as

$$\frac{-G(|\mathbf{B}_S(\mathbf{r})| + B_I)B_{rf}}{2|\mathbf{B}_S(\mathbf{r})|^2} B_{rf}^z (x\cos\delta_z - y\sin\delta_z). \quad (6)$$

From Eq. (4)-(6), we can obtain the controllable asymmetry of the rf-induced potentials with the amplitude and relative phase shift of the rf-field which is parallel to the local static field in the vicinity of the trap minima, $B_{rf}^z$ and $\delta_z$. The controllability can be understood as the result of modulating the coupling strength by changing the polarization of the rf-field. Fig. 1 (c, d) shows the asymmetric double well and ring potentials calculated with the static Ioffe-Pritchard trap and the three components rf-field (2 kHz contours) for the typical set of experimental parameters $G$ = 50 T/m, $B_I$ = 1 G, $B_{rf}$ = 1 G, $B_{rf}^z$ = 100 mG, and $\omega_{rf} = 2\pi \times 650$ kHz for $^{87}$Rb atoms in the $F = m_F$ = 2 hyperfine state.

Fig. 1 (a, b) is the schematic of the atom chip for realization of the asymmetric rf-induced potentials in our experiment. A dc current $I_S^T$ in the trap wire is used to produce an Ioffe-Pritchard trapping potential, where as rf-currents through rf-wires with $\omega_{rf}$, $I_{rf}^A$ and $I_{rf}^B$, provide the oscillating magnetic fields, $B_{rf}^A$ and $B_{rf}^B$. Static and oscillating currents through the dimple wire, $I_S^D$ and $I_{rf}^D$, is used for enhancing the longitudinal confinement as well as controlling the asymmetry of the rf-induced potentials by manipulating oscillating homogeneous field along z-direction, $B_{rf}^z$, in the vicinity of the static potential minima, respectively. For this study, we prepare a nearly pure BEC of ~5 × 10$^4$ $^{87}$Rb atoms in the $F = m_F$ = 2 hyperfine state in an elongated (along the z-direction) magnetic microtrap on the atom chip. The detailed apparatus and procedures we use for creating BEC are described elsewhere [37, 38].

For the controllable asymmetric matter-wave beam splitter, the initial BEC is positioned at a distance 135 μm from the chip surface, with radial and axial trap frequencies of 866 Hz and 26 Hz, respectively. The condensate is then split by using the three components rf magnetic field produced by currents in the chip wires. The Larmor frequency at the static trap center is 700 kHz, and $\omega_{rf}$ is fixed at a detuning of $2\pi \times 50$ kHz below this. By ramping up the rf amplitudes to the final $I_{rf} \simeq 100$ mA, over 15 ms, which is sufficient to guarantee adiabaticity during the split, the asymmetric double well is formed. Simulations predict that the separations of two minima are far enough to prevent tunnel coupling [39]. So, the asymmetry can be checked with the distribution of the atoms between the two wells. As the separations are beyond the resolution of our imaging system, in order to monitor the spatial distribution, we induced momentum on the atoms by turning off the rf-field keeping the static magnetic trap turned on (referred to as the asymmetry-checking procedure in the following) [40]. Since this procedure projects the atoms onto the bare $m_F$ states, the clouds from the two wells projected onto $m_F$ = 1 and 2 states are accelerated towards the center of the static trap while the $m_F$ = 0 atoms fall freely and the high-field seeking state atoms are rapidly expelled from the trap. After ~800 μs, that is, 3-quarters of the single harmonic trap's period, the static trap is also switched off and the clouds are separated far enough to be optically resolved with 2 ms time-of-flight [Fig. 2(a, b)].

The asymmetry of the double well is modulated with the $I_{rf}^D$. Fig. 2(c, d) shows the population ratio of the double well as a function of the final $I_{rf}^D$ in the range of our performance. The population ratios and positions of the two wells were determined by counting the numbers and by position-measurements of the atoms in $m_F$ = 1 and 2 states from each well after the asymmetry-checking procedure [41]. The positions of the wells are almost not perturbed as the asymmetry is varied over wide range, as expected from numerical calculations. Non-linearity of the dependency in the case of a vertical double well can be understood as the sagging due to gravity ($\Delta U$ = 2.1389 kHz/μm). Since the gravity effect causes a BEC to be trapped solely in the lower well, in previous studies, the position of the double well should be closer to atom chip to achieve symmetric splitting [7, 12-15]. However, in our scheme, symmetry is simply achieved with the proper choice of amplitude of $I_{rf}^D$ wherever the positions of rf-induced double well are.

A second case of interest is a ring potential on an atom chip. We were able to implement a ring potential on an atom chip by adjusting the asymmetry of the rf-induced potential. For the ring potential, the initial BEC is positioned at 102 μm directly below the trapping wire where the rf-fields from the two rf-wires are perpendicular to each other as given by Eq. (2), with radial and axial trap frequencies of 1367 Hz and 34 Hz, respectively. With $\delta = \pi/2$ and $I_{rf}^A = I_{rf}^B = I_{rf}$, the rf-induced potential forms a ring geometry for the trapped hyperfine state atoms. $\omega_{rf}$ and the ramping time are same with the values for the matter-wave beam splitter. Final $I_{rf}$ = 57.5 mA is adjusted for the diameter of the ring as ~4 μm, which is sufficiently confined to check the atom distribution by the asymmetry-checking procedure (The ring radius can be modulated by varying the amplitude and the frequency of oscillating magnetic fields). As demonstrated in [7], the trapped atoms are positioned at low bottom of the ring potential due to gravity, when $I_{rf}^D = 0$ [Fig. 3(a)]. We were able to confirm the nearly homogeneous atomic population around the ring with $I_{rf}^D$ = 1.77 mA and $\delta_z = \pi/4$ since these condition tilt the ring potential opposite the direction of gravity, as numerically calculated with actual experimental conditions. Though the compensated ring potential has inhomogeneity along the ring geometry < 1 kHz, the variation is less than the chemical potential of the trapped BEC > 2.6 kHz which is given in [31] as $\mu_c = \hbar\bar{\omega}\sqrt{2Na/\pi r_0}$ (for 3D limit), where $\bar{\omega} = \sqrt{\omega_r \omega_z}$, $a = 100 a_0$ is the scattering length (in units of the Bohr radius), $r_0$ is the radius and $N$ is the number of atoms in the ring trap. The difference between the potential variation and the chemical potential can be reduced as the variation can be reduced by decreasing the radius of the ring, while the chemical potential is a function of the controllable parameters, especially, $\omega_z$, which can be modulated with $I_S^D$ without perturbing the other parameters.

The tilt directions of the ring potential can also be controlled by the relative phase of the rf-current through the dimple wire $\delta_z$ as expected from Eq. (6) [Fig. 4(a, b)]. The experimental results are highly consistent with the numerical calculation based on the actual atom chip dimensions and $I_{rf}^D$ = 3.3 mA. The non-linear response of the relative phase is due to gravity and anisotropic coupling-strength of the rf-fields.

In this manner, the angular velocity of the tilted ring potential can be controlled with frequency difference between $I_{rf}$ and $I_{rf}^D$, since the difference acts as a continuous change of the $\delta_z$ with the phase velocity $2\pi\Delta f$. To confirm the rotation due to the frequency difference, we measure the velocity of the trapped BECs in the rotating tilted ring potential by turning off the trapping fields after 1 ms rotating trap maintenance time. With 3 ms time-of-flight, by analyzing the displacements of the untrapped atoms, the velocities were measured as a function of the frequency differences [Fig. 4(c)]. Introducing controllable rotation into the system is necessary to study the dynamics of superfluid in trapped ring BEC.

In conclusion, we have constructed a controllable asymmetric matter-wave beam splitter and a ring potential on an atom chip by dressing static magnetic trap with a three components rf-field. By controlling the vector characteristics of the rf-field, the versatile controllability on the rf-induced potentials was obtained. Asymmetry of the rf-induced double well is manipulated without discernible displacement of the each well. The controllability on the asymmetry will be a useful tool for atom interferometry since it can be used for compensating position-dependent asymmetry of matter-wave beam splitters [10, 42] or act as a phase shifter [4]. Dynamical behaviors of superfluid, such as Josephson dynamics [19-21, 43], also can be a subject with our scheme due to the precise modulation for the bias between two wells. Formation of an isotropic ring potential on an atom chip is achieved by compensating the gradient due to gravity and inhomogeneous coupling strength. In addition, position and velocity of a BEC along the ring geometry are controlled by the relative phase and the frequency difference between the rf-fields, respectively. Realizing ultracold atoms in the strongly correlated fractional quantum Hall regime has been a long-term goal of the research community [44, 45]. The smooth modifying between ring-shaped and harmonic trap, attainable very-low-energy scale of the system, and controllable versatility on the potential allowing the dynamics of the atomic system covering its angular momentum with high degree of accuracy are the most prominent characteristics for this goal [46-48].

This research was supported by a grant to the Atomic Interferometer Research Laboratory for the National Defense funded by DAPA/ADD and by Basic Science Research Program through the National Research Foundation of Korea (NRF) funded by the Ministry of Science, ICT & Future Planning (2014R1A2A2A01007460).

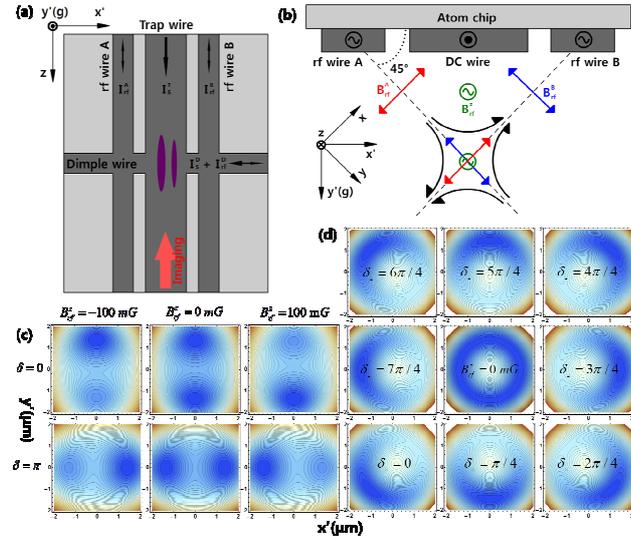

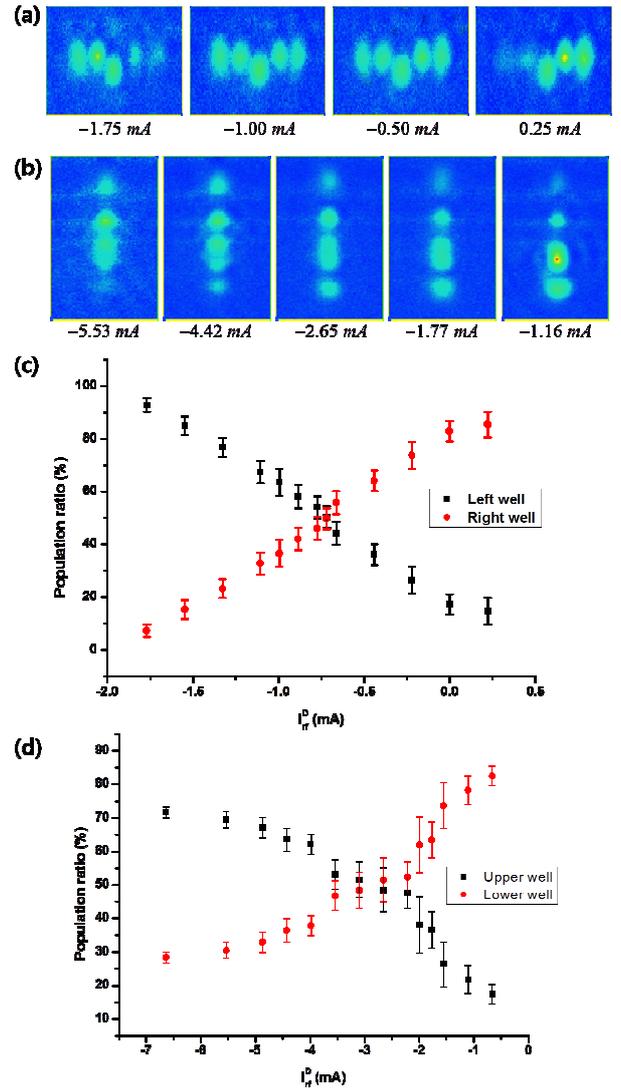

FIG. 1 (color online). (a, b) Schematic of the atom chip for asymmetric rf-induced potentials. (a) Top view of the four-wire setup. A broad (100 μm) central trap-wire carrying a dc current is used with an external bias field to produce an Ioffe-Pritchard trapping potential. Smaller (50 μm) wires on each side separated by 102 μm from the trap wire provide the oscillating magnetic fields that create the adiabatic rf-induced potentials. Running dc and rf-currents through the dimple wire is used for enhancing the longitudinal confinement and controlling the asymmetry of the rf-induced potentials by manipulating oscillating homogeneous field parallel to the quantization axis in the vicinity of the static potential minima, respectively. The atom distribution is imaged along the longitudinal axis of the trap by resonant absorption imaging. (b) Side view of the atom chip showing the relevant wires. The static trap is positioned at the two rf fields from two rf-wires are nearly perpendicular at its center. (c, d) Contour plot of the asymmetric rf-induced potentials with Ioffe-Pritchard static trap in the z = 0 plane. (c) Applying additional rf-field along z-direction can be used for controlling the asymmetry of the double well potentials in both cases, vertical ($\delta = 0$) and horizontal ($\delta = \pi$) splitting. (d) Simulation of a tilted ring trap ($\delta = \pi/2$). The direction and the degree of the tilt are controlled by the relative phase $\delta_z$ and $B_{rf}^z$, respectively.

FIG. 2 (color online). Controllable asymmetric double-well potentials on an atom chip. The asymmetry modulated with $I_{rf}^D$ which manipulating oscillating field along z-direction in the vicinity of the trap center. The positions of the wells are almost not perturbed as the asymmetry is varied over wide range. (a, b) Images of the atoms from the double wells after asymmetry-checking procedure [see text]. The clouds on left and right (top and bottom) wings are projected $m_F$ = 1 and 2 state atoms from each well (projected $m_F$ = 0 atoms positioned at the center). (c, d) Population ratio of atoms in the wells in dependence of $I_{rf}^D$. The population ratio is determined by counting the atoms in the $m_F$ = 1 and 2 states from each well. Non-linearity of the dependency in the case of vertical double well can be understood as the sagging due to gravity.

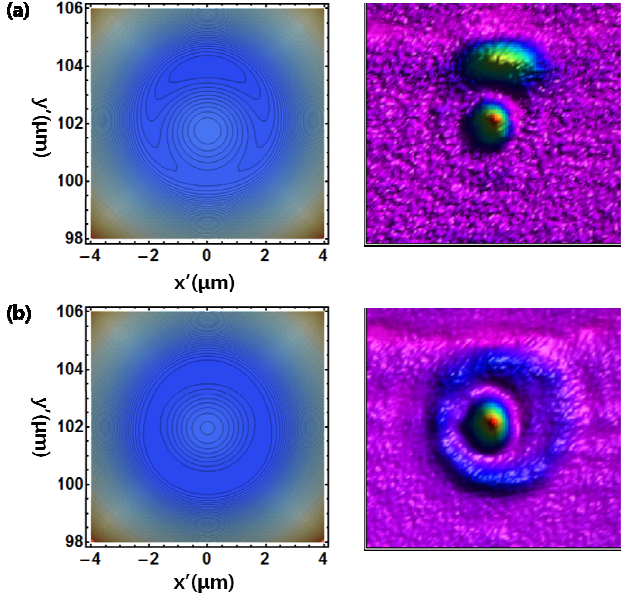

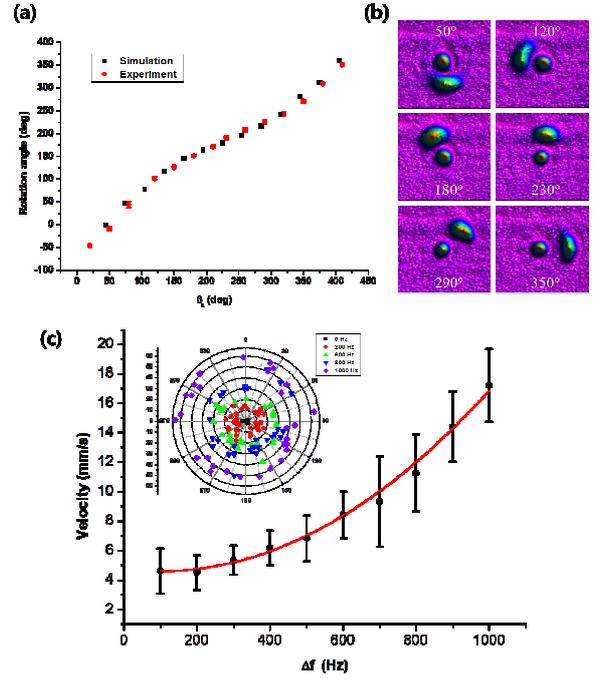

FIG. 3 (color online). Ring potential on an atom chip with $I_{rf}$ = 57.5 mA and $\delta = \pi/2$. Left - Full numerical calculation with the exact wire geometry and parameters used in the experiments, including gravity (2 kHz contour), Right - image of a BEC in a ring trap after asymmetry-checking procedure [49] (The atoms in the center of the ring are $m_F$ = 0 or high-field seeking states). (a) Anisotropic ring potential due to gravity and inhomogeneous coupling strength of the rf-field by the different distances from the rf-wires, when $I_{rf}^D$ = 0. (b) Isotropic ring potential. The anisotropy is compensated by tilt the ring potential with $I_{rf}^D$ = 1.77 mA and $\delta_z = \pi/4$.

FIG. 4 (color online). (a, b) The directions of the tilt versus the relative phase of the rf-current through the dimple wire, $\delta_z$. (a) The experimental results and expected values from the simulation with actual condition for the experiments. (b) Typical absorption images of a BEC in a tilted ring trap after asymmetry-checking procedure [49]. (c) Velocity of a BEC in a rotating tilted ring trap. The angular velocity of the tilted ring potential has been controlled with frequency difference between $I_{rf}$ and $I_{rf}^D$. To confirm that the frequency difference induces rotation on a BEC, the rotating tilted-trap turned off after 1 ms maintaining time. Onset shows the displacements of the BECs after turning off the trap and 3 ms time-of-flight.